\documentclass[10pt,twocolumn,letterpaper]{article}

\usepackage[utf8]{inputenc}
\usepackage{times}
\usepackage{epsfig}
\usepackage{graphicx}
\usepackage{amsmath}
\usepackage{amssymb}
\usepackage{hyperref}
\usepackage{color}
\usepackage[switch]{lineno}
\usepackage{authblk}
\usepackage{algorithm}
\usepackage{algpseudocode}

\def\ie{{i.e.~}}
\def\eg{{e.g.~}}
\def\cf{{c.f.~}}
\def\etal{{et~al.~}}
\newcommand{\comm}[1]{}

\newcommand\blfootnote[1]{%
  \begingroup
  \renewcommand\thefootnote{}\footnote{#1}%
  \addtocounter{footnote}{-1}%
  \endgroup
}

\begin{document}

\title{The Future of Digital Health with Federated Learning\blfootnote{Disclaimer: The opinions expressed herein are those of the authors and do not necessarily represent those of the institutions they are affiliated with, e.g. the U.S. Department of Health and Human Services or the National Institutes of Health. This is a pre-print version of \mbox{\textcolor{blue}{https://www.nature.com/articles/s41746-020-00323-1}}}} 

\author[1]{Nicola Rieke}
\author[1]{Jonny Hancox}
\author[1]{Wenqi Li}
\author[1]{Fausto Milletari}
\author[1]{Holger Roth}
\author[2,3]{Shadi Albarqouni}
\author[4]{Spyridon Bakas}
\author[5]{Mathieu N. Galtier}
\author[6]{Bennett Landman}
\author[7]{Klaus Maier-Hein}
\author[8]{Sebastien Ourselin}
\author[9]{Micah Sheller}
\author[10]{Ronald M. Summers}
\author[11,12]{Andrew Trask}
\author[1]{Daguang Xu}
\author[1]{\newline Maximilian Baust}
\author[8]{M. Jorge Cardoso}

\affil[1]{NVIDIA}
%\affil[2]{Center for Biomedical Image Computing and Analytics (CBICA), University of Pennsylvania, Philadelphia, USA}
%\affil[2]{University of Pennsylvania}
\affil[2]{Technical University of Munich (TUM), Munich, Germany}
%\affil[10]{Technical University of Munich}
\affil[3]{Computer Vision Lab (CVL), ETH Zürich, Switzerland}
\affil[4]{University of Pennsylvania (UPenn) Philadelphia, USA}
\affil[5]{Owkin, Paris, France}
%\affil[4]{Departments of Radiology and Radiological Sciences, Vanderbilt University Medical Center, Nashville, USA}
\affil[6]{Vanderbilt University Medical Center, Nashville, USA}
%affil[4]{Vanderbilt University Medical Center}
\affil[7]{German Cancer Research Center (DKFZ), Heidelberg, Germany}
%\affil[5]{Division of Medical Image Computing, German Cancer Research Center (DKFZ), Heidelberg, Germany}
%\affil[5]{German Cancer Research Center (DKFZ)}
\affil[8]{King’s College London (KCL), London, UK}
%\affil[6]{Biomedical Engineering and Imaging Sciences, King’s College London, London, UK}
%\affil[6]{King’s College London}
\affil[9]{Intel Corporation, USA}
%\affil[8]{Imaging Biomarkers and Computer-Aided Diagnosis Lab, Radiology and Imaging Sciences, National Institutes of Health (NIH) Clinical Center}
\affil[10]{Clinical Center, National Institutes of Health (NIH), Bethesda, Maryland, USA}
\affil[11]{OpenMined}
\affil[12]{University of Oxford, Oxford, UK}

\renewcommand\Authands{ and }

%\date{}
\date{\vspace{-5ex}}

\maketitle

\begin{abstract}

\textit{
Data-driven Machine Learning has emerged as a promising approach for building accurate and robust statistical models from medical data, which is collected in huge volumes by modern healthcare systems.
Existing medical data is not fully exploited by ML primarily because it sits in data silos and privacy concerns restrict access to this data. However, without access to sufficient data, ML will be prevented from reaching its full potential and, ultimately, from making the transition from research to clinical practice. %\newline
This paper considers key factors contributing to this issue,  explores how Federated Learning (FL) may provide a solution for the future of digital health and highlights the challenges and considerations that need to be addressed.
}
\end{abstract}

\section{Introduction}
Research on artificial intelligence (AI) has enabled a variety of significant breakthroughs over the course of the last two decades.
In digital healthcare, the introduction of powerful Machine Learning-based and particularly Deep Learning-based models~\cite{lecun2015deep} has led to disruptive innovations in radiology, pathology, genomics and many other fields.
In order to capture the complexity of these applications, modern Deep Learning (DL) models feature a large number (e.g. millions) of parameters that are learned from and validated on medical datasets.
Sufficiently large corpora of curated data are thus required in order to obtain models that yield clinical-grade accuracy, whilst being safe, fair, equitable and generalising well to unseen data~\cite{chartrand2017deep,de2018clinically,sun2017revisiting}.

For example, training an automatic tumour detector and diagnostic tool in a supervised way requires a large annotated database that encompasses the full spectrum of possible anatomies, pathological patterns and types of input data. Data like this is hard to obtain and curate.
One of the main difficulties is that unlike other data, which may be shared and copied rather freely, health data is highly sensitive, subject to regulation and cannot be used for research without appropriate patient consent and ethical approval~\cite{van2014systematic}.
Even if data anonymisation is sometimes proposed as a way to bypass these limitations, it is now well-understood that removing metadata such as patient name or date of birth is often not enough to preserve privacy~\cite{rocher2019estimating}. Imaging data suffers from the same issue - it is possible to reconstruct a patient's face from three-dimensional imaging data, such as computed tomography (CT) or magnetic resonance imaging (MRI). Also the human brain itself has been shown to be as unique as a fingerprint~\cite{yeh2016quantifying}, where subject identity, age and gender can be predicted and revealed~\cite{wachinger2015brainprint}. 
Another reason why data sharing is not systematic in healthcare is that medical data are potentially highly valuable and costly to acquire. Collecting, curating and maintaining a quality dataset takes considerable time and effort. These datasets may have a significant business value and so are not given away lightly. In practice, openly sharing medical data is often restricted by data collectors themselves, who need fine-grained control over the access to the data they have gathered.

Federated Learning (FL)~\cite{mcmahan2017communication,li2019federated, yang2019federated} is a learning paradigm that seeks to address the problem of data governance and privacy by training algorithms collaboratively without exchanging the underlying datasets. 
The approach was originally developed in a different domain, but it recently gained traction for healthcare applications because it neatly addresses the problems that usually exist when trying to aggregate medical data.
Applied to digital health this means that FL enables insights to be gained collaboratively across institutions, e.g. in the form of a global or consensus model, without sharing the patient data. 
In particular, the strength of FL is that sensitive training data does not need to be moved beyond the firewalls of the institutions in which they reside.
Instead, the Machine Learning (ML) process occurs locally at each participating institution and only model characteristics (\eg parameters, gradients etc.) are exchanged.
Once training has been completed, the trained consensus model benefits from the knowledge accumulated across all institutions.
Recent research has shown that this approach can achieve a performance that is comparable to a scenario where the data was co-located in a data lake and superior to the models that only see isolated single-institutional data~\cite{li2019privacy, sheller2018multi}. 

For this reason, we believe that a successful implementation of FL holds significant potential for enabling precision medicine at large scale.
The scalability with respect to patient numbers included for model training would facilitate models that yield unbiased decisions, optimally reflect an individual's physiology, and are sensitive to rare diseases in a way that is respectful of governance and privacy concerns.
Whilst FL still requires rigorous technical consideration to ensure that the algorithm is proceeding optimally without compromising safety or patient privacy, it does have the potential to overcome the limitations of current approaches that require a single pool of centralised data. 
\newline

The aim of this paper is to provide context and detail for the community regarding the benefits and impact of FL for medical applications (Section~\ref{sec:Motivation}) as well as highlighting key considerations and challenges of implementing FL in the context of digital health (Section~\ref{sec:Considerations}).
The medical FL use-case is inherently different from other domains, e.g. in terms of number of participants and data diversity, and while recent surveys investigate the research advances and open questions of FL~\cite{kairouz2019advances,yang2019federated,xu2019federated}, we focus on what it actually means for digital health and what is needed to enable it. 
We envision a federated future for digital health and hope to inspire and raise awareness with this article for the community. 

\begin{figure*}[!t]
\begin{center}
   \includegraphics[width=\textwidth]{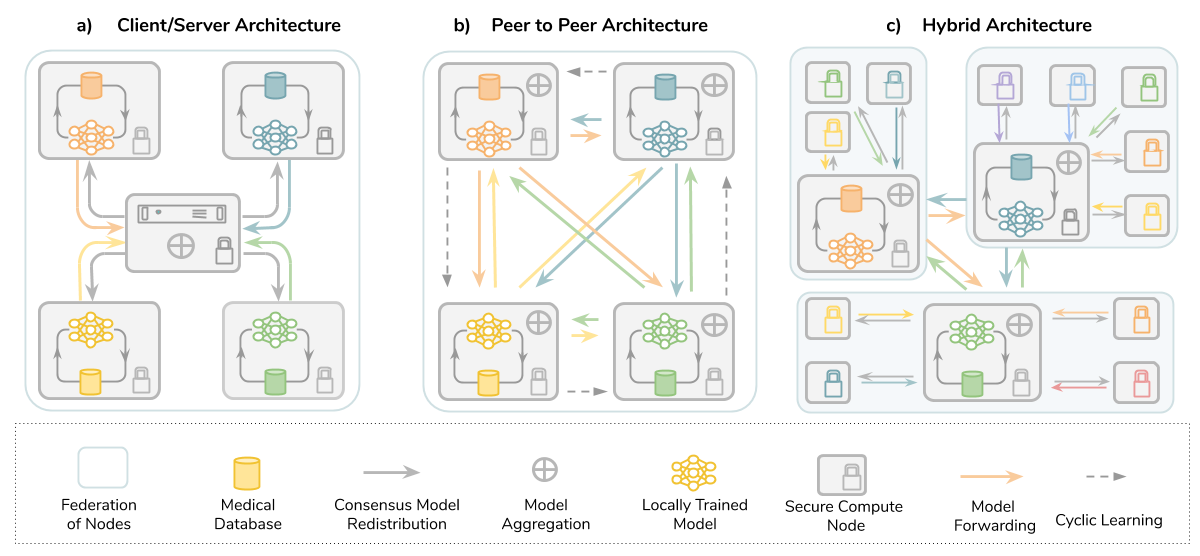}
\end{center}
\caption{\textbf{Federated Learning Communication Architectures} - (a) Client-Server architecture via Hub and Spokes: A Parameter Server distributes the model and each node trains a local model for several iterations, after which the updated models are returned to the Parameter Server for aggregation. This consensus model is then redistributed for subsequent iterations. (b) Decentralised Architecture via peer-to-peer: Rather than using a Parameter Server, each node broadcasts its locally trained model to some or all of its peers and each node does its own aggregation. (c) Hybrid Architecture: Federations can be composed into a hierarchy of hubs and spokes, which might represent regions, health authorities or countries.
}
    \label{fig:fl_system}
\end{figure*}

\section{Data-driven Medicine Requires Federated Efforts}
\label{sec:Motivation}

ML and especially DL is becoming the \textit{de facto} knowledge discovery approach in many industries, but successfully implementing data-driven applications requires that models are trained and evaluated on sufficiently large and diverse datasets. 
These medical datasets are difficult to curate (Section~\ref{sec:Data}). FL offers a way to counteract this data dilemma and its associated governance and privacy concerns by enabling collaborative learning without centralising the data (Section ~\ref{sec:Promise}). This learning paradigm, however, requires consideration from and offers benefits to the various stakeholders of the healthcare environment (Section~\ref{sec:Stakeholders}). 
All these points will be discussed in this section.

\subsection{The Reliance on Data}
\label{sec:Data}
Data-driven approaches rely on datasets that truly represent the underlying data distribution of the problem to be solved.
Whilst the importance of comprehensive and encompassing databases is a well-known requirement to ensure generalisability, state-of-the-art algorithms are usually evaluated on carefully curated datasets, often originating from a small number of sources - if not a single source. This implies major challenges: pockets of isolated data can introduce sample bias in which demographic (e.g. gender, age etc.) or technical imbalances (e.g. acquisition protocol, equipment manufacturer) skew the predictions, adversely affecting the accuracy of prediction for certain groups or sites.  

The need for sufficiently large databases for AI training has spawned many initiatives seeking to pool data from multiple institutions.
Large initiatives have so far primarily focused on the idea of creating data lakes. These data lakes have been built with the aim of leveraging either the commercial value of the data, as exemplified by IBM's Merge Healthcare acquisition~\cite{IBM:Online}, or as a resource for economic growth and scientific progress, with examples such as NHS Scotland's National Safe Haven~\cite{NHS:Online}, the French Health Data Hub~\cite{cuggia2019french} and Health Data Research UK~\cite{HDRUK:Online}.
Substantial, albeit smaller, initiatives have also made data available to the general community such as \textit{The Human Connectome}~\cite{sporns2005human}, UK Biobank~\cite{sudlow2015uk}, \textit{The Cancer Imaging Archive} (TCIA)~\cite{clark2013cancer}, NIH CXR8~\cite{wang2017chestx}, NIH DeepLesion~\cite{yan2018deep},  \textit{The Cancer Genome Atlas} (TCGA)~\cite{tomczak2015cancer}, the \textit{Alzheimer's disease neuroimaging initiative} (ADNI)~\cite{jack2008alzheimer}, or as part of medical grand challenges\footnote{https://grand-challenge.org/} such as the  CAMELYON challenge~\cite{litjens20181399}, the multimodal brain tumor image segmentation benchmark (BRATS)~\cite{menze2014multimodal} or the Medical Segmentation Decathlon~\cite{simpson2019large}. Public data is usually task- or disease-specific and often released with varying degrees of license restrictions, sometimes limiting its exploitation.
Regardless of the approach, the availability of such data has the potential to catalyse scientific advances, stimulate technology start-ups and deliver improvements in healthcare.

Centralising or releasing data, however, poses not only regulatory and legal challenges related to ethics, privacy and data protection, but also technical ones - safely anonymising, controlling access, and transferring healthcare data is a non-trivial, and often impossible, task. As an example, anonymised data from the electronic health record can appear innocuous and GDPR/PHI compliant, but just a few data elements may allow for patient reidentification~\cite{rocher2019estimating}. The same applies to genomic data and medical images, with their high-dimensional nature making them as unique as one's fingerprint~\cite{yeh2016quantifying}. Therefore, unless the anonymisation process destroys the fidelity of the data, likely rendering it useless, patient reidentification or information leakage cannot be ruled out.
Gated access, in which only approved users may access specific subsets of data, is often proposed as a putative solution to this issue. However, not only does this severely limit data availability, it is only practical for cases in which the consent granted by the data owners or patients is unconditional, since recalling data from those who may have had access to the data is practically unenforceable. 

\subsection{The Promise of Federated Efforts}
\label{sec:Promise}

\begin{figure*}[!t]
\begin{center}
   \includegraphics[width=\textwidth]{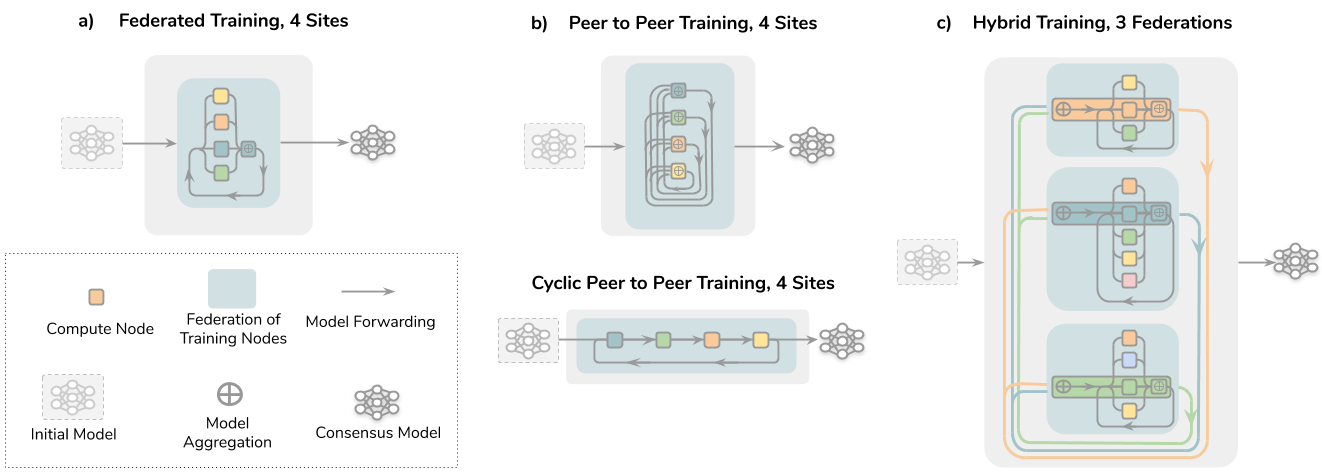}
\end{center}
\caption{\textbf{Compute Plan} a) Federated Training: Standard approach to federation in which each of the nodes train in parallel and submit their model updates for aggregation every few epochs. The aggregation may happen on one of the training nodes or a separate Parameter Server node, which would then redistribute the consensus model. 
b) Peer to Peer Training: Nodes broadcast their model updates to one or more nodes in the federation and each does its own aggregation. Cyclic Training happens when model updates are passed to a single neighbour one or more times, round-robin style.
c) Hybrid Training: Federations, perhaps in remote geographies, can be composed into a hierarchy and use different communication/aggregation strategies at each tier.  In the illustrated case, three federations of varying size periodically share their models using a peer to peer approach. The consensus model is then redistributed to each federation and each node therein.}
    \label{fig:fl_computeplan}
\end{figure*}

The promise of FL is simple - to address privacy and governance challenges by allowing algorithms to learn from non co-located data. In a FL setting, each data controller not only defines their own governance processes and associated privacy considerations, but also, by not allowing data to move or to be copied, controls data access and the possibility to revoke it. So the potential of FL is to provide controlled, indirect access to large and comprehensive datasets needed for the development of ML algorithms, whilst respecting patient privacy and data governance. It should be noted that this includes both the training as well as the validation phase of the development. In this way, FL could create new opportunities, e.g. by allowing large-scale validation across the globe directly in the institutions, and enable novel research on, for example, rare diseases, where the incident rates are low and it is unlikely that a single institution has a dataset that is sufficient for ML approaches.
Moving the to-be-trained model to the data instead of collecting the data in a central location has another major advantage: the high-dimensional, storage-intense medical data does not have to be duplicated from local institutions in a centralised pool and duplicated again by every user that uses this data for local model training. In a FL setup, only the model is transferred to the local institutions and can scale naturally with a potentially growing global dataset without replicating the data or multiplying the data storage requirements.

Some of the promises of FL are implicit: a certain degree of privacy is provided since other FL participants never directly access the data from other institutions and only receive the resulting model parameters that are aggregated over several participants. 
And in a Client-Server Architecture (see Figure~\ref{fig:fl_system}), in which a federated server manages the aggregation and distribution, the participating institutions can even remain unknown to each other. However, it has been shown that the models themselves can, under certain conditions, memorise information~\cite{shokri2017membership,sablayrolles2019white,zhang2016understanding,carlini2019secret}. Therefore the FL setup can be further enhanced by privacy protections using mechanisms such as differential privacy~\cite{abadi2016deep,shokri2015privacy} or learning from encrypted data (\cf Sec.~\ref{sec:Considerations}). And FL techniques are still a very active area of research~\cite{kairouz2019advances}.

All in all, a successful implementation of FL will represent a paradigm shift from centralised data warehouses or lakes, with a significant impact on the various stakeholders in the healthcare domain.
\subsection{Impact on Stakeholders}
\label{sec:Stakeholders}
If FL is indeed the answer to the challenge of healthcare ML at scale, then it is important to understand who the various stakeholders are in a FL ecosystem and what they have to consider in order to benefit from it.

\paragraph{Clinicians} are usually exposed to only a certain sub-group of the population based on the location and demographic environment of the hospital or practice they are working in.
Therefore, their decisions might be based on biased assumptions about the probability of certain diseases or their interconnection. 
By using ML-based systems, \eg as a second reader, they can augment their own expertise with expert knowledge from other institutions, ensuring a consistency of diagnosis not attainable today.
Whilst this promise is generally true for any ML-based system, systems trained in a federated fashion are potentially able to yield even less biased decisions and higher sensitivity to rare cases as they are likely to have seen a more complete picture of the data distribution.
In order to be an active part of or to benefit from the federation, however, demands some up-front effort such as compliance with agreements e.g. regarding the data structure, annotation and report protocol, which is necessary to ensure that the information is presented to collaborators in a commonly understood format. 

\paragraph{Patients} are usually relying on local hospitals and practices. 
Establishing FL on a global scale could ensure higher quality of clinical decisions regardless of the location of the deployed system.
For example, patients who need medical attention in remote areas could benefit from the same high-quality ML-aided diagnosis that are available in hospitals with a large number of cases.
The same advantage applies to patients suffering from rare, or geographically uncommon, diseases, who are likely to have better outcomes if faster and more accurate diagnoses can be made.
FL may also lower the hurdle for becoming a data donor, since patients can be reassured that the data remains with the institution and data access can be revoked.

\paragraph{Hospitals and Practices} can remain in full control and possession of their patient data with complete traceability of how the data is accessed.
They can precisely control the purpose for which a given data sample is going to be used, limiting the risk of misuse when they work with third parties.
However, participating in federated efforts will require investment in on-premise computing infrastructure or private-cloud service provision. 
The amount of necessary compute capabilities depends of course on whether a site is only participating in evaluation and testing efforts or also in training efforts.
Even relatively small institutions can participate, since enough of them will generate a valuable corpus and they will still benefit from collective models generated.
One of the drawbacks is that FL strongly relies on the standardisation and homogenisation of the data formats so that predictive models can be trained and evaluated seamlessly.
This involves significant standardisation efforts from data managers.
 
\paragraph{Researchers and AI Developers} who want to develop and evaluate novel algorithms stand to benefit from access to a potentially vast collection of real-world data.
This will especially impact smaller research labs and start-ups, who would be able to directly develop their applications on healthcare data without the need to curate their own datasets. 
By introducing federated efforts, precious resources can be directed towards solving clinical needs and associated technical problems rather than relying on the limited supply of open datasets.
At the same time, it will be necessary to conduct research on algorithmic strategies for federated training, \eg how to combine models or updates efficiently, how to be robust to distribution shifts, etc., as highlighted in the technical survey papers~\cite{kairouz2019advances,yang2019federated,xu2019federated}.
And a FL-based development implies that the researcher or AI developer cannot investigate or visualise all of the data on which the model is trained. It is for example not possible to look at an individual failure case to understand why the current model performs poorly on it. 

\paragraph{Healthcare Providers} in many countries are affected by the ongoing paradigm shift from volume-based, \ie fee-for-service-based, to value-based healthcare.
A value-based reimbursement structure is in turn strongly connected to the successful establishment of precision medicine.
This is not about promoting more expensive individualised therapies but instead about achieving better outcomes sooner through more focused treatment, thereby reducing the costs for providers. By way of example, with sufficient data, ML approaches can learn to recognise cancer-subtypes or genotypic traits from radiology images that could indicate certain therapies and discount others.
So, by providing exposure to large amounts of data, FL has the potential to increase the accuracy and robustness of healthcare AI, whilst reducing costs and improving patient outcomes, and is therefore vital to precision medicine.

\paragraph{Manufacturers} of healthcare software and hardware could benefit from federated efforts and infrastructures for FL as well, since combining the learning from many devices and applications, without revealing anything patient-specific can facilitate the continuous improvement of ML-based systems. This potentially opens up a new source of data and revenue to manufacturers. However, hospitals may require significant upgrades to local compute, data storage, networking capabilities and associated software to enable such a use-case. 
Note, however, that this change could be quite disruptive: FL could eventually impact the business models of providers, practices, hospitals and manufacturers affecting patient data ownership; and the regulatory frameworks surrounding continual and FL approaches are still under development.

\section{Technical Considerations}
\label{sec:Considerations}
FL is perhaps best-known from the work of Kone{\v{c}}n{\`y} \etal  ~\cite{konevcny2016distributed}, but various other definitions have been proposed in literature~\cite{kairouz2019advances, yang2019federated,xu2019federated,mcmahan2017communication}.
These approaches can be realised via different communication architectures (see Figure~\ref{fig:fl_system}) and respective compute plans (see Figure~\ref{fig:fl_computeplan}).
The main goal of FL, however, remains the same: to combine knowledge learned from non co-located data, that resides within the participating entities, into a global model. 
Whereas the initial application field mostly comprised mobile devices, participating entities in the case of healthcare could be institutions storing the data, \eg hospitals, or medical devices itself, \eg a CT scanner or even low-powered devices that are able to run computations locally.
It is important to understand that this domain-shift to the medical field implies different conditions and requirements. 
For example, in the case of the federated mobile device application, potentially millions of participants could contribute, but it would be impossible to have the same scale of consortium in terms of participating hospitals. On the other hand medical institutions may rely on more sophisticated and powerful compute infrastructure with stable connectivity. 
Another aspect is that the variation in terms of data type and defined tasks as well as acquisition protocol and standardisation in healthcare is significantly higher than pictures and messages seen in other domains.
The participating entities have to agree on a collaboration protocol and the high-dimensional medical data, which is predominant in the field of digital health, poses challenges by requiring models with huge numbers of parameters. This may become an issue in scenarios where the available bandwidth for communication between participants is limited, since the model has to be transferred frequently.
And even though data is never shared during FL, considerations about the security of the connections between sites as well as mitigation of data leakage risks through model parameters are necessary. 
In this section, we will discuss more in detail what FL is and how it differs from similar techniques as well as highlighting the key challenges and technical considerations that arise when applying FL in digital health.

\subsection{Federated Learning Definition}
FL is a learning paradigm in which multiple parties train collaboratively without the need to exchange or centralise datasets.
Although various training strategies have been implemented to address specific tasks, a general formulation of FL can be formalised as follows:
Let $\mathcal{L}$ denote a global loss function obtained via a weighted combination of $K$ local losses $\{\mathcal{L}_k\}_{k=1}^{K}$, computed from private data $X_{k}$ residing at the individual involved parties:
\begin{align}
\min_{\phi}\mathcal{L}(X; \phi) \quad \text{with} \quad  \mathcal{L}(X; \phi)=\sum_{k=1}^{K}w_{k}\;\mathcal{L}_{k}(X_{k}; \phi),
\label{eq:formalism}
\end{align}
where $w_k>0$ denote the respective weight coefficients.
It is important to note that the data $X_{k}$ is never shared among parties and remains private throughout learning.

In practice, each participant typically obtains and refines the global consensus model by running a few rounds of optimisation on their local data and then shares the updated parameters with its peers, either directly or via a parameter server. 
The more rounds of local training are performed without sharing updates or synchronisation, the less it is guaranteed that the actual procedure is minimising the equation (\ref{eq:formalism})~\cite{mcmahan2017communication, kairouz2019advances}.
The actual process used for aggregating parameters commonly depends on the FL network topology, as FL nodes might be segregated into sub-networks due to geographical or legal constraints (see Figure~\ref{fig:fl_system}). Aggregation strategies can rely on a single aggregating node (hub and spokes models), or on multiple nodes without any centralisation. An example of this is peer-to-peer FL, where connections exist between all or a subset of the participants and model updates are shared only between directly-connected sites~\cite{roy2019braintorrent,lalitha2019peer}.
An example of centralised FL aggregation with a client-server architecture is given in Algorithm~\ref{alg}.
Note that aggregation strategies do not necessarily require information about the full model update; clients might choose to share only a subset of the model parameters for the sake of reducing communication overhead of redundant information, ensure better privacy preservation~\cite{li2019federated} or to produce multi-task learning algorithms having only part of their parameters learned in a federated manner.

A unifying framework enabling various training schemes may disentangle compute resources (data and servers) from the \textit{compute plan}, as depicted in Figure~\ref{fig:fl_computeplan}. The latter defines the trajectory of a model across several partners, to be trained and evaluated on specific datasets. 

For more details regarding state-of-the-art of FL techniques, such as aggregation methods, optimisation or model compression, we refer the reader to the overview by Kairouz \etal~\cite{kairouz2019advances}.

\begin{algorithm}[t]
\small{
  \caption{Example of a FL algorithm~\cite{li2019privacy} in a client-server architecture with aggregation via \texttt{FedAvg}~\cite{mcmahan2017learning}. }\label{alg}
  \begin{algorithmic}[1]
    \Require{num\_federated\_rounds $T$}
    %\Require{number of local training iterations ($N_k^{(local)}$) }
    \Procedure{Aggregating}{}
      \State{Initialise global model: $W^{(0)}$}
      \For{$t \gets 1\cdots T$}
        \For{$client\ k \gets 1\cdots K$}  \Comment{ \textit{Run in parallel}}
        \State{Send $W^{(t-1)}$ to client $k$}
        \State{Receive model updates and number of local training iterations $(\Delta W_k^{(t-1)}, N_k)$ from client's local training with $\mathcal{L}_{k}(X_{k}; W^{(t-1)})$} 
        
        \EndFor
        \State{$W^{(t)}\gets W^{(t-1)}+\frac{1}{\sum_k{N_k}}\sum_k{(N_k\cdot \Delta W_k^{(t-1)})}$}
      \EndFor
      \State \Return $W^{(t)}$
    \EndProcedure
  \end{algorithmic}}
\end{algorithm}

\subsection{Relation to Similar Strategies}
FL is rooted in older forms of collaborative learning where models are shared or compute is distributed~\cite{sheller2018multi,chang2018distributed}.  
Transfer Learning, for example, is a well-established approach of model-sharing that makes it possible to tackle problems with deep neural networks that have millions of parameters, despite the lack of extensive, local datasets that are required for training from scratch: 
a model is first trained on a large dataset and then further optimised on the actual target data. The dataset used for the initial training does not necessarily come from the same domain or even the same type of data source as the target dataset. This type of transfer learning has shown better performance~\cite{shin2016deep,tajbakhsh2016convolutional} when compared to strategies where the model had been trained from scratch on the target data only, especially when the target dataset is comparably small. 
It should be noted that similar to a FL setup, the data is not necessarily co-located in this approach. For Transfer Learning, however, the models are usually shared acyclically, e.g. using a pre-trained model to fine-tune it on another task, without contributing to a collective knowledge-gain. And, unfortunately, deep learning models tend to "forget"~\cite{mccloskey1989catastrophic,goodfellow2013empirical}. Therefore after a few training iterations on the target dataset the initial information contained in the model is lost~\cite{li2017learning}.
To adopt this approach into a form of collaborative learning in a FL setup with continuous learning from different institutions, the participants can share their model with a peer-to-peer architecture in a "round-robin" or parallel fashion and train in turn on their local data.
This yields better results when the goal is to learn from diverse datasets. A client-server architecture in this scenario enables learning on multi-party data at the same time~\cite{sheller2018multi}, possibly even without forgetting~\cite{shoham2019overcoming}.

There are also other collaborative learning strategies~\cite{song2018collaborative,ruder2017overview} such as ensembling, a statistical strategy of combining multiple independently trained models or predictions into a consensus, or multi-task learning, a strategy to leverage shared representations for related tasks. 
These strategies are independent of the concept of FL, and can be used in combination with it. 

The second characteristic of FL - to distribute the compute - has been well studied in recent years~\cite{jin2016scale,goyal2017accurate,ben2019demystifying}.
Nowadays, the training of the large-scale models is often executed on multiple devices and even multiple nodes~\cite{ben2019demystifying}. In this way, the task can be parallelised and enables fast training, such as training a neural network on the extensive dataset of the ImageNet project in 1 hour~\cite{goyal2017accurate} or even in less than 80 seconds~\cite{yamazaki2019yet}.
It should be noted that in these scenarios, the training is realised in a cluster environment, with centralised data and fast network communication. 
So, distributing the compute for training on several nodes is feasible and FL may benefit from the advances in this area. 
Compared to these approaches, however, FL comes with a significant communication and synchronisation cost.
In the FL setup, the compute resources are not as closely connected as in a cluster and every exchange may introduce a significant latency. Therefore, it may not be suitable to synchronise after every batch, but to continue local training for several iterations before aggregation.

We refer the reader to the survey by Xu \emph{et al.}~\cite{xu2019federated} for an overview of the evolution of Federated Learning and the different concepts in the broader sense.

\subsection{Challenges and Considerations}
Despite the advantages of FL, there are challenges that need to be taken into account when establishing federated training efforts.
In this section, we discuss five key aspects of FL that are of particular interest in the context of its application to digital health. 

\subsubsection{Privacy and Security}
In healthcare, we work with highly sensitive data that must be protected accordingly. Therefore, some of the key considerations are the trade-offs, strategies and remaining risks regarding the privacy-preserving potential of FL.  

\textbf{Privacy Vs. Performance.} Although one of the main purposes of FL is to protect privacy by sharing model updates rather than data, FL does not solve all potential privacy issues and - similar to ML algorithms in general - will always carry some risks. 
Strict regulations and data governance policies make any leakage, or perceived risk of leakage, of private information unacceptable. These regulations may even differ between federations and a catch-all solution will likely never exist. Consequently, it is important that potential adopters of FL are aware of potential risks and state-of-the-art options for mitigating them. 
Privacy-preserving techniques for FL offer levels of protection that exceed today's current commercially available ML models~\cite{kairouz2019advances}.
However, there is a trade-off in terms of performance and these techniques may affect for example the accuracy of the final model~\cite{li2019federated}. 
Furthermore future techniques and/or ancillary data could be used to compromise a model previously considered to be low-risk.

\textbf{Level of Trust}. Broadly speaking, participating parties can enter two types of FL collaboration:

\textit{Trusted} - for FL consortia in which all parties are considered trustworthy and are bound by an enforceable collaboration agreement, we can eliminate many of the more nefarious motivations, such as deliberate attempts to extract sensitive information or to intentionally corrupt the model. This reduces the need for sophisticated counter-measures, falling back to the principles of standard collaborative research. 

\textit{Non-trusted} - in FL systems that operate on larger scales, it is impractical to establish an enforceable collaborative agreement that can guarantee that all of the parties are acting benignly. Some may deliberately try to degrade performance, bring the system down or extract information from other parties.
In such an environment, security strategies will be required to mitigate these risks such as, encryption of model submissions, secure authentication of all parties, traceability of actions, differential privacy, verification systems, execution integrity, model confidentiality and protections against adversarial attacks.

\textbf{Information leakage.}
By definition, FL systems sidestep the need to share healthcare data among participating institutions. 
However, the shared information may still indirectly expose private data used for local training, for example by model inversion~\cite{wu_p3sgd:_2019} of the model updates, the gradients themselves~\cite{zhu2019deep} or adversarial attacks~\cite{wang2019beyond,hitaj_deep_2017}. FL is different from traditional training insofar as the training process is exposed to multiple parties. 
As a result, the risk of leakage via reverse-engineering increases if adversaries can observe model changes over time, observe specific model updates (\ie a single institution’s update), or manipulate the model (\eg induce additional memorisation by others through gradient-ascent-style attacks). Countermeasures, such as limiting the granularity of the shared model updates and to add specific noise to ensure differential privacy~\cite{abadi2016deep,li2019privacy,li2020multi} may be needed and is still an active area of research~\cite{kairouz2019advances}. 

\subsubsection{Data heterogeneity}
Medical data is particularly diverse - not only in terms of type, dimensionality and characteristics of medical data in general but also within a defined medical task, due to factors like acquisition protocol, brand of the medical device or local demographics. This poses a challenge for FL algorithms and strategies: one of the core assumptions of many current approaches is that the data is independent and identically distributed (IID) across the participants.
Initial results indicate that FL training on medical non-IID data is possible, even if the data is not uniformly distributed across the institutions~\cite{li2019privacy,sheller2018multi}.
In general however, strategies such as \textit{FedAvg} \cite{mcmahan2017communication} are prone to fail under these conditions~\cite{li2018federated,mcmahan2016communication,zhao2018federated}, in part defeating the very purpose of collaborative learning strategies. Research addressing this problem includes for example \emph{FedProx}~\cite{li2018federated} and part-data-sharing strategy~\cite{zhao2018federated}.
Another challenge is that data heterogeneity may lead to a situation in which the global solution may not be the optimal final local solution.
The definition of model training optimality should therefore be agreed by all participants before training.

\subsubsection{Traceability and accountability}
As per all safety-critical applications, the reproducibility of a system is important for FL in healthcare. In contrast to training on centralised data, FL involves running multi-party computations in environments that exhibit complexities in terms of hardware, software and networks.
The traceability requirement should be fulfilled to ensure that system events, data access history and training configuration changes, such as hyperparameter tuning, can be traced during the training processes.
Traceability can also be used to log the training history of a model and, in particular, to avoid the training dataset overlapping with the test dataset.
In particular in non-trusted federations, traceability and accountability processes running in require execution integrity.
After the training process reaches the mutually agreed model optimality criteria, it may also be helpful to measure the amount of contribution from each participant, such as computational resources consumed, quality of the local training data used for local training etc.  The measurements could then be used to determine relevant compensation and establish a revenue model among the participants \cite{ghorbani2019data}.

One implication of FL is that researchers are not able to investigate images upon which models are being trained. 
So, although each site will have access to its own raw data, federations may decide to provide some sort of secure intra-node viewing facility to cater for this need or perhaps even some utility for explainability and interpretability of the global model. However, the issue of interpretability within DL is still an open research question. 

\subsubsection{System architecture}
Unlike running large-scale FL amongst consumer devices, healthcare institutional participants are often equipped with better computational resources and
reliable and higher throughput networks. These enable for example training of larger models with larger numbers of local training steps and sharing more model information between nodes.
This unique characteristic of FL in healthcare consequently brings opportunities as well as challenges such as (1) how to ensure data integrity when communicating (e.g. creating redundant nodes); (2) how to design secure encryption methods to take advantage of the computational resources; (3) how to design appropriate node schedulers and make use of the distributed computational devices to reduce idle time.

The administration of such a federation can be realised in different ways, each of which come with advantages and disadvantages. In high-trust situations, training may operate via some sort of 'honest broker' system, in which a trusted third party acts as the intermediary and facilitates access to data.
This setup requires an independent entity to control the overall system which may not always be desirable, since it could involve an additional cost and procedural viscosity, but does have the advantage that the precise internal mechanisms can be abstracted away from the clients, making the system more agile and simpler to update.
In a peer-to-peer system each site interacts directly with some or all of the other participants. In other words, there is no gatekeeper function, all protocols must be agreed up-front, which requires significant agreement efforts, and changes must be made in a synchronised fashion by all parties to avoid problems.
And in a trustless-based architecture the platform operator may be cryptographically locked into being honest which creates significant computation overheads whilst securing the protocol.

\subsubsection{Initiatives and consortia}
Future efforts to apply artificial intelligence to healthcare tasks may strongly depend on collaborative strategies between multiple institutions rather than large centralised databases belonging to only one hospital or research laboratory. The ability to leverage FL to capture and integrate knowledge acquired and maintained by different institutions provides an opportunity to capture larger data variability and analyse patients across different demographics. Moreover, FL is an opportunity to incorporate multi-expert annotation and multi-centre data acquired with different instruments and techniques. This collaborative effort requires, however, various agreements including definitions of scope, aim and technology which, since it is still novel, may incorporate several unknowns. In this context, large-scale initiatives such as the \emph{MELLODDY} project~\footnote{http://www.imi.europa.eu/projects-results/project-factsheets/melloddy}, the HealthChain project~\footnote{https://www.substra.ai/en/healthchain-project}, the Trustworthy Federated Data Analytics (TFDA) and the German Cancer Consortium's Joint Imaging Platform (JIP)~\footnote{https://www.dkfz.de/en/datascience/platforms-initiatives.html} represent pioneering efforts to set the standards for safe, fair and innovative collaboration in healthcare research.

\section{Conclusion}
ML, and particularly DL, has led to a wide range of innovations in the area of digital healthcare.
As all ML methods benefit greatly from the ability to access data that approximates the true global distribution, FL is a promising approach to obtain powerful, accurate, safe, robust and unbiased models. By enabling multiple parties to train collaboratively without the need to exchange or centralise datasets, FL neatly addresses issues related to egress of sensitive medical data.  
As a consequence, it may open novel research and business avenues and has the potential to improve patient care globally.
In this article, we have discussed the benefits and the considerations pertinent to FL within the healthcare field. 
Not all technical questions have been answered yet and FL will certainly be an active research area throughout the next decade~\cite{kairouz2019advances}. Despite this, we truly believe that its potential impact on precision medicine and ultimately improving medical care is very promising.

\section{Acknowledgement}

%This research was supported by the UK Research and Innovation London Medical Imaging \& Artificial Intelligence Centre for Value Based Healthcare, by the Wellcome Trust, by the Intramural Research Program of the National Institutes of Health Clinical Center, as well as by the Helmholtz Initiative and Networking Fund (project "Trustworthy Federated Data Analytics")
This work was supported by the UK Research and Innovation London Medical Imaging \& Artificial Intelligence Centre for Value-Based Healthcare, by the Wellcome/EPSRC Centre for Medical Engineering (WT203148/Z/16/Z), by the Wellcome Flagship Programme (WT213038/Z/18/Z), by the Intramural Research Programme of the National Institutes of Health (NIH) Clinical Center, by the National Cancer Institute of the NIH under award number U01CA242871, by the National Institute of Neurological Disorders and Stroke of the NIH under award number R01NS042645, as well as by the Helmholtz Initiative and Networking Fund (project “Trustworthy Federated Data Analytics”) and the PRIME programme of the German Academic Exchange Service (DAAD) with funds from the German Federal Ministry of Education and Research (BMBF). The content and opinions expressed in this publication is solely the responsibility of the authors and do not necessarily represent those of the institutions they are affiliated with, e.g., the U.S. Department of Health and Human Services or the National Institutes of Health. 

%Financial Disclosure: Author RMS receives royalties from iCAD, ScanMed, Philips, and Ping An. His lab has received research support from Ping An and NVIDIA. Author SA is supported by the PRIME programme of the German Academic Exchange Service (DAAD) with funds from the German Federal Ministry of Education and Research (BMBF). Author SB is supported by the National Institutes of Health (NIH). Author MNG is supported by the HealthChain (BPIFrance) and Melloddy (IMI2) projects.

Competing Interests / Financial Disclosure: R.M.S. receives royalties from iCAD, ScanMed, Philips, Translation Holdings and Ping An. His lab has received research support from Ping An and NVIDIA. S.B. is supported by the National Institutes of Health (NIH). M.N.G. is supported by the HealthChain (BPIFrance) and Melloddy (IMI2) projects. A.T. is an employee of Google’s DeepMind. S.O. and M.J.C. are founders and shareholders of Brainminer, llc. The other authors declare no competing interests.

\bibliographystyle{IEEEtran}
\bibliography{ref}

\end{document}